\documentclass[aps,pra,showpacs,superscriptaddress,twocolumn]{revtex4}
\usepackage{amsmath}
\usepackage{amsfonts}
\usepackage{graphicx}
\usepackage{float}
\begin{document}

\title{High-order harmonics as induced by a quantized field: \\ a phase-space picture}
\author{\'{A}kos Gombk\"{o}t\H{o}}
\affiliation{Department of Theoretical Physics, University of Szeged, Tisza Lajos k\"{o}r%
\'{u}t 84, H-6720 Szeged, Hungary}
\author{S\'{a}ndor Varr\'{o}}
\affiliation{Wigner Research Centre for Physics, Konkoly-Thege M. \'{u}t 29-33, H-1121 Budapest, Hungary}
\affiliation{ELI-ALPS, ELI-HU Non-profit Ltd., Dugonics t\'{e}r 13, H-6720 Szeged, Hungary}%
\author{P\'{e}ter Mati}
\affiliation{ELI-ALPS, ELI-HU Non-profit Ltd., Dugonics t\'{e}r 13, H-6720 Szeged, Hungary}%
\author{P\'{e}ter F\"{o}ldi}
\affiliation{Department of Theoretical Physics, University of Szeged, Tisza Lajos k\"{o}r%
\'{u}t 84, H-6720 Szeged, Hungary}
\affiliation{ELI-ALPS, ELI-HU Non-profit Ltd., Dugonics t\'{e}r 13, H-6720 Szeged, Hungary}%

\begin{abstract}
The interaction of matter with a quantized electromagnetic mode is considered. Representing a strong exciting field, the mode is assumed to contain a large number of photons. As a result, the material response is highly nonlinear: the completely quantized description results in generation of high harmonics. In order to understand the essence of the physical processes that are involved, we consider a finite dimensional model for the material system. Using an appropriate description in phase space, this approach leads to a transparent picture showing that the interaction splits the initial, exciting coherent state into parts, and the rapid change of the populations of these parts (that are coherent states themselves) results in the generation of high-order harmonics as secondary radiation. The method we use is an application of the discrete lattice of coherent states that was introduced by J.~von Neumann.
\end{abstract}

\maketitle

\section{Introduction}
The importance of high-order harmonic generation (HHG) is unquestionable in recent development of high-field laser science. The process itself provides fundamental information on nonlinear light-matter interaction (see, e.g., \cite{HLScKHKKH15,NGWBScGR16}), while, on the other hand, it has practical applications, e.g., in the production of ultrashort bursts of electromagnetic radiation \cite{FT1992,HKSRMBCHDK2001,PTBMABMA2001}. The appearance of HH frequencies has been detected for the case of target materials ranging from gas samples \cite{McPherson87,Ferray88} via surfaces \cite{QTMDMGA2006,Vincenti2014} till wide bandgap solids \cite{GDiCSADiMR11,Ghim19}.

Traditionally, the theoretical description of the phenomenon of HHG relies on the semiclassical approximation, i.e., on the assumption that the exciting radiation can be treated as a classical, time-dependent field \cite{Cor93,Lew94}. In fact, usually the HH modes themselves are also considered to be classical, despite the fact that they contain orders of magnitude less photons than the excitation. A quantum optical analysis of the harmonic modes was considered in Ref.~\cite{GC16}, where the time-dependent populations of these modes together with the corresponding photon statistics were given.

However, there are experimental implications showing that a completely quantized description is required for the full understanding of the physical processes that are responsible for HHG. As reported in Refs.~\cite{Tsatrafyllis2017,T19}, by measuring the photon statistics of a strong, mid-IR pulse after the interaction with a gas \cite{Tsatrafyllis2017} or a solid \cite{T19} sample, one can identify fingerprints of the harmonics. In other words, the back-action of the material system on the exciting field is observable -- {\em on the quantized level}.

Considering theory, a general description of a free charged particle interacting with a quantized mode has already been given for both the relativistic \cite{BV81b} and the nonrelativistic \cite{BV81a} cases. In fact, in Ref.~\cite{BV81b} the very first non-perturbative treatment of HHG in the nonlinear Compton process has been given, in the frame of the fully quantized description. Transitions between Volkov states have also been used for the description of HHG process in atoms \cite{GSE98,CCL00}.  The theoretical models developed in  Ref.~\cite{Gonos16} and Ref.~\cite{T19}  for gases and crystalline samples, support the experimental findings reported in Refs.~\cite{Tsatrafyllis2017,T19}. However, there are still a number of open questions, and besides experimental results, the detailed physical understanding of the interaction between quantized light and matter in the high-intensity regime requires further theoretical investigations as well.

For the sake of clarity, in the following we consider an approach that has already been proven to be very useful for the description of traditional quantum optical problems. By using the Jaynes-Cummings-Paul model (without rotating wave approximation), we show that a very transparent interpretation of the process of the HHG can be given on the phase space of the exciting mode. Let us note that this model can directly be related to HHG in quantum wells \cite{H94}, where only a finite number of states get excited, or solid state HHG using the velocity gauge, where all transitions are "vertical", i.e., the dynamics of states with different $\mathbf{k}$ are independent (see, e.g., \cite{KK12,FP2017}). Moreover, since any numerical approach unavoidably uses a finite dimensional Hilbert-space, our approach --that is based on von Neumann lattice coherent sates \cite{N32} -- points towards the development of a general, efficient scheme for calculating the dynamics of strong quantized fields that interact with matter.

In the current paper, first, in Sec.~\ref{modelsec}, we present the model to be used, and show how the Hamiltonian in the strong-field approximation can be diagonalized. In Sec.~\ref{latticesec} we expand the initial state on the basis of the von Neumann lattice coherent sates and show how Wigner functions can be calculated using this expansion. The analysis of the process of HHG by the aid of the corresponding Wigner functions is performed in Sec.~\ref{mainsec}. HHG spectra are presented in Sec.~\ref{spectrumsec}. Possible generalizations of our model are discussed in Sec.~\ref{generalizationsec}, and the conclusions are drawn in Sec.~\ref{finalsec}.

\section{Model and strong-field approximation}
\label{modelsec}
In order to see the essence of the interaction of strong, quantized fields and matter, let us consider the case of a two-level atom ($a$) and a single mode ($f$):
\begin{equation}
H=H_a+H_f+H_{af}=\frac{\hbar}{2}\omega_0\sigma_z +\hbar\omega a^\dagger a + \hbar\Omega\sigma_x(a+a^\dagger).
\label{HamS}
\end{equation}
Here the usual creation and annihilation operators (with $[a,a^\dagger]=1$) and the Pauli matrices appear.
The eigenstates of the atomic Hamiltonian will be denoted by $|e\rangle$ and $|g\rangle$, i.e., $H_a|e\rangle=\hbar\omega_0/2 |e\rangle,$ $H_a|g\rangle=-\hbar\omega_0/2 |g\rangle.$ We will also use $|+\rangle$ and $|-\rangle,$ for which $\sigma_x|\pm\rangle=\pm|\pm\rangle.$ The state of the field will be described using the photon number eigensates, $H_f|n\rangle=\hbar\omega n |n\rangle.$ We use the convention for the order of factors in tensor products as "atom, field," e.g. $|e\rangle |n\rangle.$ In the following we focus on far offresonant excitation, for definiteness we assume $\omega_0/\omega=2.2,$ which corresponds to the ratio of the bandgap in ZnO \cite{GDiCSADiMR11} and the photon energy for commonly used sources at $800$ nm wavelength. (Note that this choice has no qualitative effect on the results.)

 Eq.~(\ref{HamS}) above describes the well-known Jaynes-Cummings-Paul model without using rotating-wave approximation (RWA). (Note that RWA does not allow the appearance of high order harmonics.) Unlike the case when RWA can be applied (see e.g.~Ref.~\cite{MS91}), there is no known general solution to this model. However, resolvent techniques \cite{W72,T00} were applied for the analysis of the problem, and the resulting formulae allowed numerical calculations that showed that RWA does not work precisely enough for strong coupling, not even in the resonant case of $\omega_0=\omega$ \cite{T00}. A comparison of the solutions based on continued fractions to the case with RWA was performed also in Ref.~\cite{S72}. The eigenvalues of the infinite matrix of $H$ in the photon number eigenstate basis were discussed in Ref.~\cite{MNS04}, while multiphoton generalizations of the model were investigated e.g., in Ref.~\cite{NLL99}. In Ref.~\cite{ZZ88}, a path integral approach to the non-RWA description of the problem [as given by an interaction picture version of Eq.~(\ref{HamS})] was presented.

 It is closely related to our present work that by choosing appropriate time dependent parameters, the number of the dynamical equations can become finite \cite{DK73,D80}. Davydov Ans\"{a}tze thus simplify the calculation of the time evolution to a large extent, see e.g. \cite{WG18}. Additionally, methods that use multi-Davydov Ans\"{a}tze \cite{ZHZ15} are similar to the case of solving the dynamics on the von-Neumann lattice (to be described hereafter). In this context, the relevance of using the von-Neumann lattice, which is a very universal concept, stems from the possibility of generalizing the results.

\subsection{Time evolution in the absence of the atomic Hamiltonian}

We are to solve the dynamics induced by $H$ for a given initial state, which, in order to serve as a model for HHG, corresponds to a high mean photon number. As we shall see in the following, the sum of the second and third terms in the Hamiltonian $H$ can be diagonalized, and $H_a$ can be taken into account as an action additional to the strong field approximation described by
\begin{equation}
\tilde{H}=H - H_a = \hbar\omega a^\dagger a + \hbar\Omega\sigma_x(a+a^\dagger).
\end{equation}

As a first step for the diagonalization of $\tilde{H},$ let us define the generalized displacement operator
\begin{equation}
\tilde{D}(\gamma)=e^{\sigma_x\left(\gamma a^\dagger - \gamma^* a \right)},
\end{equation}
which is unitary, $\tilde{D}^{-1}(\gamma)=\tilde{D}^{\dagger}(\gamma)=\tilde{D}(-\gamma).$ As we shall see, it is sufficient for now to consider transformations with real valued $\gamma$ parameters only. Since
\begin{equation}
\left[ \sigma_x\gamma \left( a - a^\dagger \right) ,a\right]=\sigma_x\gamma, \ \ \ \left[ \sigma_x\gamma \left( a - a^\dagger \right) ,a^\dagger\right]=\sigma_x\gamma,
\end{equation}
and because the right hand sides commute with the exponent in $\tilde{D},$ we can use the identity $e^{A}Be^{-A}=B+[A,B]$ to obtain
\begin{equation}
\tilde{D}^\dagger(\gamma)a \tilde{D}(\gamma)=a+\gamma\sigma_x, \ \ \ \tilde{D}^\dagger(\gamma)a^\dagger \tilde{D}(\gamma)=a^\dagger+\gamma\sigma_x,
\label{DaD}
\end{equation}
and
\begin{align}
\tilde{D}^\dagger(\gamma) a^\dagger a \tilde{D}(\gamma)&=\tilde{D}^\dagger(\gamma)a \tilde{D}(\gamma) \tilde{D}^\dagger(\gamma) a^\dagger \tilde{D}(\gamma)\nonumber \\
&=a^\dagger a + \gamma \sigma_x\left(a + a^\dagger \right) +\gamma^2 \sigma_x^2.
\label{DaaD}
\end{align}
By the aid of Eqs.~(\ref{DaD}) and (\ref{DaaD}), the transformation of $\tilde{H}$ reads:
\begin{align}
\tilde{D}^\dagger(\gamma)\tilde{H}\tilde{D}(\gamma)&=\hbar \omega \left[ a^\dagger a + \gamma \sigma_x\left(a + a^\dagger \right) +\gamma^2 \sigma_x^2 \right]\nonumber \\ &+ \hbar\Omega\sigma_x\left( a+a^\dagger + 2\gamma \sigma_x \right).
\end{align}
By collecting the coefficients of the products $a \sigma_x$ and $a^\dagger \sigma_x,$ we can see that the choice $\gamma=-\Omega/\omega$ reduces the transformed Hamiltonian to \begin{equation}
\tilde{D}^\dagger\left(\frac{-\Omega}{\omega}\right)\tilde{H}\tilde{D}\left(\frac{-\Omega}{\omega}\right)=\hbar \omega \left(a^\dagger a - \frac{\Omega^2}{\omega^2} \sigma_x^2 \right).
\label{transformedH}
\end{equation}
Note that so far we intentionally did not use the identity $\sigma_x^2=1.$ This means that the results are general in the sense that they are valid for arbitrary coupling operator that can replace $\sigma_x,$ which will be useful later on (see Sec.~\ref{generalizationsec}). However, in order the see the most transparent results of the model, from now on  we use the special properties of the coupling operator $\sigma_x.$ This results in a transformed Hamiltonian (\ref{transformedH}) which is proportional to the identity on the atomic subspace.

This means that
\begin{equation}
\tilde{D}^\dagger\left(\frac{-\Omega}{\omega}\right)\tilde{H}\tilde{D}\left(\frac{-\Omega}{\omega}\right)|\phi\rangle|n\rangle=E_n |\phi\rangle|n\rangle,
\label{eigen1}
\end{equation}
where $E_n=\hbar \omega \left(n - \frac{\Omega^2}{\omega^2}\right),$ and surprisingly, in the current case, $|\phi\rangle$ can be an arbitrary atomic state. Moreover, since the constant $- \frac{\Omega^2}{\omega^2}$ in $E_n$ will only lead to an irrelevant global phase in the time evolution, we will omit it in the following. (Formally, this only means the redefinition of the zero level of the energy.)   Multiplying (\ref{eigen1}) by $\tilde{D}$ from the left, we obtain
\begin{equation}
\tilde{H}\left[\tilde{D}\left(\frac{-\Omega}{\omega}\right)|\phi\rangle|n\rangle\right]=E_n \left[\tilde{D}\left(\frac{-\Omega}{\omega}\right) |\phi\rangle|n\rangle\right],
\label{eigen2}
\end{equation}
i.e., the states $\tilde{D}(\frac{-\Omega}{\omega})|\phi\rangle|n\rangle$ are eigenstates of $\tilde{H}.$

\bigskip

Let us analyze the time evolution induced by $\tilde{H}$ alone. For the sake of simplicity, $|\Psi\rangle(t)$ will denote the solution of $i\hbar\frac{\partial}{\partial t}|\Psi\rangle=\tilde{H}|\Psi\rangle,$ and we use $\gamma=-\Omega/\omega.$ It is convenient to consider states that are eigenstates of $\sigma_x,$ when $\tilde{D}(\gamma)$ acts as the usual displacement operator $D(\gamma)$:
\begin{align}
\tilde{D}(\gamma)|\pm\rangle|n\rangle&=D(\pm \gamma) |\pm\rangle |n\rangle \\ \nonumber
&=|\pm\rangle e^{\pm\left(\gamma a^\dagger - \gamma^* a \right)}|n\rangle=|\pm\rangle |n,\pm\gamma\rangle,
\end{align}
where displaced photon number eigenstates \cite{OKKB90} appear on the right hand side. Any initial state can be expanded as
\begin{equation}
|\Psi\rangle (0) =\sum_{n} b_n^+ |+\rangle |n,\gamma\rangle + b_n^- |-\rangle |n,-\gamma\rangle,
\label{general0}
\end{equation}
leading to the time evolution
\begin{equation}
|\Psi\rangle (t) =\sum_{n} \left(b_n^+ |+\rangle |n,\gamma\rangle + b_n^- |-\rangle |n,-\gamma\rangle\right)e^{-i n \omega t}. \label{general}
\end{equation}
(Let us recall that the irrelevant global phase factor $\exp(i \frac{\Omega^2}{\omega^2}t )$ is ignored.)

As an important application, we calculate the time evolution of $|\Psi\rangle (0)=|+\rangle|\alpha\rangle=D(\alpha)|+\rangle|0\rangle.$ Note that the coherent state $|\alpha\rangle=\exp(-|\alpha|^2/2)\sum \alpha^n/\sqrt{n!}|n\rangle$ with a large magnitude complex label $\alpha$ is the most appropriate description for a laser mode with a high mean photon number ($\langle n \rangle=|\alpha|^2).$  By using the expansion above, Eq.~(\ref{general}) leads to
\begin{equation}
|\pm\rangle|\alpha\rangle(t)=e^{i\delta_\pm(t)}|+\rangle|\alpha_\pm(t)\rangle, \label{coh1t}
\end{equation}
with $\alpha_+(t)=\pm\gamma+(\alpha \mp\gamma)e^{-i\omega t}$ and $\delta_\pm(t)=\pm\gamma \mathrm{Im}\  [\alpha-(\alpha\mp\gamma)e^{-i\omega t}].$ (Recall that $\mathrm{Im} \gamma=0.$) For clarity, e.g., $|+\rangle|\alpha\rangle(t)$ above denotes the $\tilde{H}$-induced time evolution of the initial state $|+\rangle|\alpha\rangle,$ which turns out to be $\exp {i\delta_+(t)}$ times the tensor product of $|+\rangle$ and a coherent state, whose time dependent index is given by $\alpha_+(t).$

\begin{figure}[htb]
\includegraphics[width=7.9cm]{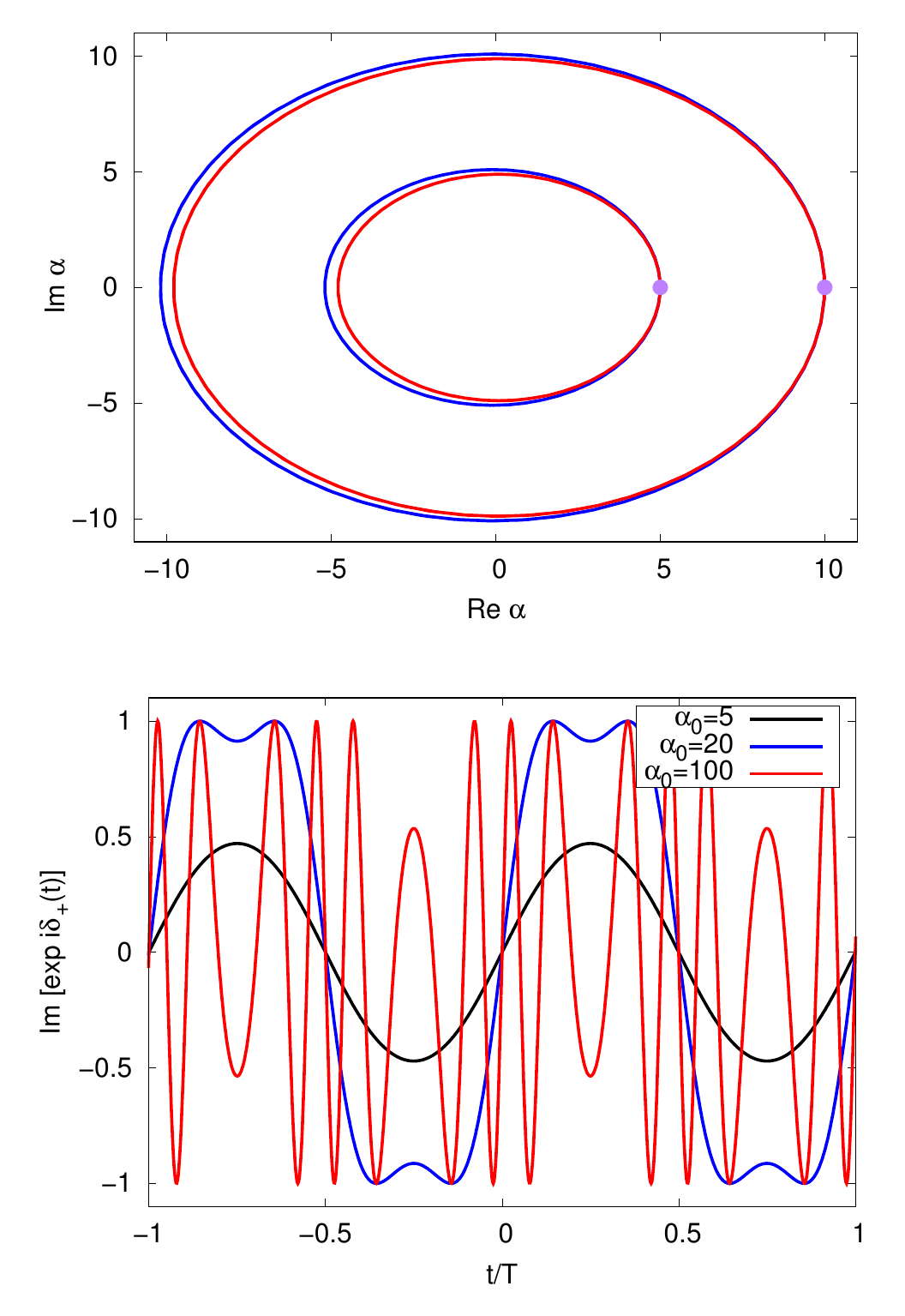}
\caption{The time evolution of coherent states as induced by $\tilde{H}.$ Top panel: the labels $\alpha_\pm(t)$ of $|\pm\rangle|\alpha_\pm(t)\rangle$ as curves on the complex plane (red circle: $-$ sign, blue circle: $+$ sign) for two different values of $\alpha_0$ that are denoted by the purple dots. Bottom panel: the imaginary part of the phase $\exp[i\delta_+(t)]$ during two optical cycles. $\gamma=0.1$ for both panels.}
\label{cohfig}
\end{figure}

 Clearly, $\alpha_\pm(0)=\alpha$ and the exponential prefactors in Eq.~(\ref{coh1t}) reduce to unity at $t=0.$ Similarly, for integer $n$, $|\pm\rangle|\alpha\rangle(t=nT)=|\pm\rangle|\alpha\rangle(0),$ where
$T=\omega/2\pi$ is the optical cycle-time.  This means that the time evolutions of the coherent states are periodic when $H_a$ is omitted. Visually, as it is shown by the top panel of Fig.~\ref{cohfig}, the time dependent indices $\alpha_\pm(t),$ as a curves on the complex plane, describe two circles. It is also interesting to observe how high harmonics appear in the dynamics of the phases $\exp[i\delta_\pm(t)]$: when we increase the mean photon number ($|\alpha|^2$), the phases show more and more complex behavior (see the bottom panel of Fig.~\ref{cohfig}). However, if we assume, as usually, that it is the expectation value of the dipole moment operator ($\propto \sigma_x$ in our case) that is the source of the HH radiation, there are no observable harmonics in the absence of $H_a,$ since $[\tilde{H},\sigma_x]=0,$ thus the expectation value of the dipole moment is constant.
By superposition, the solutions (\ref{coh1t}) allow us to calculate the $\tilde{H}$-induced time evolution of the most plausible initial state, i.e., when the atom is in its ground state:
\begin{equation}
|\Psi\rangle(0)=|g\rangle|\alpha_0\rangle=\frac{1}{\sqrt{2}}\left(|+\rangle+|-\rangle\right)|\alpha_0\rangle.
\label{cohini}
\end{equation}
In this case, in the top panel of Fig.~\ref{cohfig}, the purple full circle represent the initial state, which periodically splits into the coherent superposition of two parts that are visualized by the red and blue curves.

\section {von Neumann lattice coherent states and Wigner functions}
\label{latticesec}
\subsection{Dynamics on the von Neumann lattice}
Since $[\tilde{H}, H_a]\neq 0,$ there is no system of common eigenstates for these operators. Let us use the results of the previous section and describe the dynamics by the aid of the eigenstates of $\tilde{H}$. Using photon number eigenstate expansion, a general solution of the time dependent Schr\"{o}dinger equation $|\Psi\rangle(t)$ for $i\hbar\frac{\partial}{\partial t}|\Psi\rangle=\tilde{H}|\Psi\rangle$ is given by (\ref{general}), where the complex coefficients $b_m^\pm$ are constant. When we use the complete Hamiltonian, the presence of $H_a$ results in the time dependence of these coefficients. Specifically, using $H_a|\pm\rangle=\hbar \omega_0 /2 |\mp\rangle,$ we obtain:
\begin{align}
i\hbar \dot{b}_n^+&=\frac{\hbar \omega_0}{2}\sum_m b_m^- \langle n| D(-2\gamma) | m\rangle e^{-i\omega(m-n)t} \nonumber \\
i\hbar \dot{b}_n^-&=\frac{\hbar \omega_0}{2}\sum_m b_m^+ \langle n| D(2\gamma) | m\rangle e^{-i\omega(m-n)t},
\label{ndyn}
\end{align}
i.e., the time evolution is not diagonal in the index $n,$ the time derivative of $b_n^+$ ($b_n^-$) involves infinitely many coefficients $b_m^-$ ($b_m^+$). This means a severe technical difficulty, despite the fact that the matrix elements on the right hand side have analytic expressions in terms of Laguerre polynomials \cite{OKKB90}. As we shall see, a finite number of coherent states (indexed by a subset of the von Neumann lattice) is a convenient choice for a basis the elements of which are not orthogonal, but practically transform among themselves under the action of $H_a$ on the timescale of HHG.

\bigskip

Coherent states are known to form an overcomplete basis on the Hilbert space of a single mode, i.e., they obey the closure relation, but not the entire complex plane as the index of the states is needed for the expansion of an arbitrary state (see e.g., Ref.~\cite{JDA93} for the case of a circle).  According to J.~von Neumann \cite{N32}, it is sufficient to use a discrete subset $\{|\alpha^{(mn)}\rangle,\mathrm{Re}\ \alpha^{(nm)}=m\sqrt{\pi},\mathrm{Im}\ \alpha^{(nm)}=n\sqrt{\pi}\}$ with $n$ and $m$ being integers. (For the proof of completeness, see Refs.~\cite{B71,P71}.) That is, any state can be written as $|\phi\rangle=\sum_{mn}c_{mn} |\alpha^{(nm)}\rangle$
in an unambiguous way. This lattice was used in Ref.~\cite{TW80} to transform the operator equations for the quantized modes into c-number equations.
For an application to the case of HHG, see Ref.~\cite{VS}.

In the following we demonstrate that the basis of von Neumann lattice coherent states is also very convenient for the calculation of the complete dynamics, as well as for the visualization of the results on phase space.

\begin{figure}[htb]
\includegraphics[width=7.9cm]{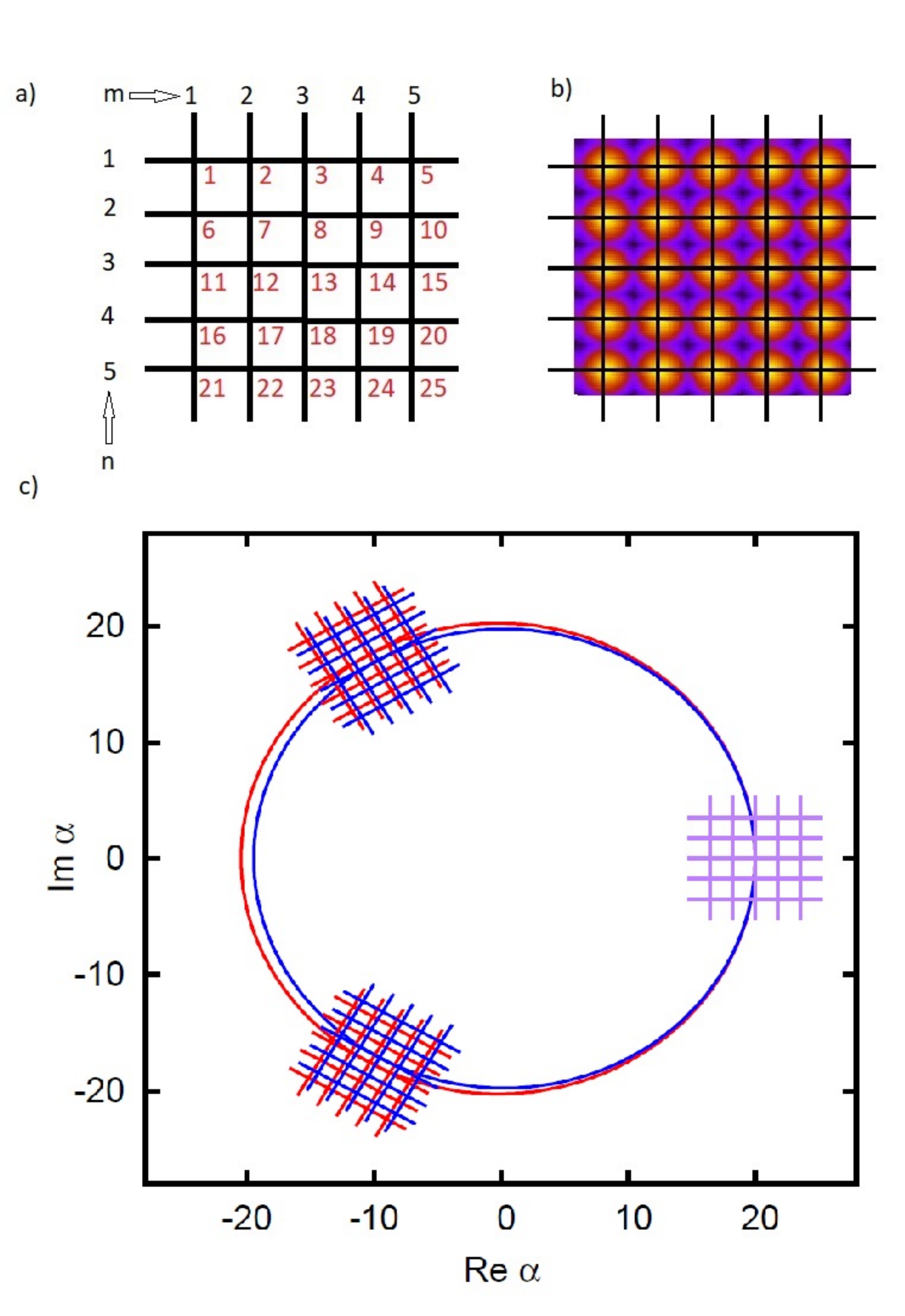}
\caption{von Neumann-lattice coherent states. Panel a): a finite set of lattice points with two different methods for generating the indices. Panel b): Wigner functions of coherent states centered at the same lattice points. Panel c): visualization of the time evolution of a finite basis set $\{|\pm\rangle |\alpha^{(k)}\rangle(t),k=1,\ldots,25\}.$}
\label{Wfig}
\end{figure}

Similarly to the case of Eq.~(\ref{cohini}), let us focus on the realistic initial state $|g\rangle|\alpha_0\rangle$
and use a finite subset of the von Neumann lattice around $\alpha_0$ to expand the initial state.  If $n_0$ and $m_0$ are indices for which $\left|\alpha^{(m_0 n_0)}-\alpha_0 \right|$ is minimal (i.e., we found the integer indices $(m_0,n_0)$ for which the corresponding von Neumann lattice point $\alpha^{(m_0 n_0)}$ is the closest to $\alpha_0$), then a grid of lattice points with $m=-N+m_0, \ldots, m_0, \ldots, N+m_0,$ $n=-N+n_0, \ldots, n_0, \ldots, N+n_0$ define an approximate basis. It is convenient to switch to a single index $k$ (e.g., by starting from one of the corners of the lattice points and keeping row-continuous order, see Fig.~\ref{Wfig}a), where $n_0=m_0=0$ for the sake of simplicity), and use the set $\{|+\rangle|\alpha^{(k)}\rangle, |-\rangle|\alpha^{(k)}\rangle\}_{k=1}^{(2N+1)\times(2N+1)}$ for the expansion of the initial state:
\begin{equation}
|\Psi\rangle(0)=|g\rangle|\alpha_0\rangle=\sum_k c^+_k |+\rangle |\alpha^{(k)}\rangle  + c^-_k|-\rangle |\alpha^{(k)}\rangle.
\label{initial}
\end{equation}
Clearly, for an exact expansion, all the coherent states of the von Neumann lattice are needed. However, as it will be underlined by a numerical example in Sec.~\ref{mainsec}, already relatively small lattices ($N\approx 5$) provide a precision that is sufficient for most practical purposes.

In order  to determine the complex coefficients $c_k^{\pm},$  we have to take into account that the overlap $\mathcal{N}_{ij}=\langle \alpha^{(i)}|\alpha^{(j)}\rangle=\exp(\alpha_i^*\alpha_j)\exp[-(|\alpha_i|^2+|\alpha_j|^2)/2]$ is not zero. If we fix the number of the lattice points (i.e., $N$), we obtain:
\begin{equation}
c_i^\pm=\sum_j  \mathcal{N}^{-1}_{ij} \langle\pm|\langle\alpha^{(j)}| \Psi\rangle(0),
\label{ccoeff}
\end{equation}
where the inverse of the $(2N+1)\times(2N+1)$ overlap matrix appears on the right hand side.

It is important that the time evolution of all the basis states is known under the action of $\tilde{H},$ see Eq.~(\ref{coh1t}).  This allows us to calculate the complete dynamics (as induced by $H=\tilde{H}+H_a$) by letting the coefficients $c_k^{\pm}$ time dependent:
\begin{equation}
|\Psi\rangle(t)=\sum_k c^+_k(t) |+\rangle |\alpha^{(k)}\rangle(t)  + c^-_k(t) |-\rangle |\alpha^{(k)}\rangle(t).
\label{timedep}
\end{equation}
Now it is worth defining the following four overlap matrices
\begin{align}
\mathcal{N}_{jm}^{\pm\pm}(t)=&\langle\alpha^{(j)}_\pm(t)| \alpha^{(m)}_\pm(t)\rangle \\
\times & \exp i\left\{\delta_\pm[\alpha^{(m)}_\pm(t)] - \delta_\pm[\alpha^{(j)}_\pm(t)]\right\},
\nonumber
\end{align}
which are all equal to $\mathcal{N}$ at $t=0.$ Note that here the upper indices, $+$ or $-,$ correspond to the lower ones they are situated right above, e.g., $\mathcal{N}_{jm}^{+-}$ is proportional to $\langle\alpha^{(j)}_+| \alpha^{(m)}_-\rangle.$ Using this notation, Eq.~(\ref{ccoeff}) is valid also in the time dependent case if we replace the matrix $\mathcal{N}$ by $\mathcal{N}^{++}(t)$ and $\mathcal{N}^{--}(t)$ for $c_i^+(t)$ and $c_i^-(t),$ respectively.

The dynamical equations for $c_k^\pm(t)$ are given by:
\begin{align}
i\hbar \sum_k \mathcal{N}_{mk}^{++}\frac{d}{d t} c_k^+(t) =& \sum_k  \mathcal{N}_{mk}^{++}(t) c_k^+(t) \langle +|H_a|+\rangle  \nonumber \\
+& \sum_k \mathcal{N}_{mk}^{+-}(t) c_k^-(t) \langle +|H_a|-\rangle.
\label{basisdyn1}
\end{align}
Recalling that the expectation value of $H_a$ vanishes in the states $|\pm\rangle$ [see before Eq.~(\ref{ndyn})], and using the inverse of the overlap matrices, we obtain:
\begin{equation}
i\hbar \frac{d}{d t} c_j^+(t) = \sum_{km} \left(\mathcal{N}^{++}\right)^{-1}_{jm}(t) \mathcal{N}_{mk}^{+-}(t) c_k^-(t) \langle +|H_a|-\rangle.
\label{coeffdyn1}
\end{equation}
Similarly:
\begin{equation}
i\hbar \frac{d}{d t} c_j^- = \sum_{km} \left(\mathcal{N}^{--}\right)^{-1}_{jm}(t) \mathcal{N}_{mk}^{-+}(t) c_k^+(t) \langle -|H_a|+\rangle.
\label{coeffdyn2}
\end{equation}
Note that -- apart from time instants that are integer multiples of optical cycle time $T$ -- the matrix products $(\mathcal{N}^{++})^{-1} \mathcal{N}^{+-}$ and $(\mathcal{N}^{--})^{-1} \mathcal{N}^{-+}$ are close, but not equal to unity. This shows how the action of $H_a$ slowly mixes different von Neumann lattice coherent states.
\bigskip

The results above can be directly turned into a numerical procedure for solving the time-dependent Schr\"{o}dinger equation using the von Neumann lattice. First we have to determine the initial part of the lattice that will serve as an approximate basis. [Clearly, this choice depends on the index $\alpha_0$ of the initial state, see Eq.~(\ref{initial}).] Then, at each time instant, $\{|+\rangle|\alpha^{(k)}\rangle(t), |-\rangle|\alpha^{(k)}\rangle\}_{k=1}^{(2N+1)\times(2N+1)(t)}$ have to be determined [see Eqs.~(\ref{coh1t})], and also the overlap matrices $\mathcal{N}^{\pm\pm}(t)$ should be updated. These calculations are completely analytic. According to our numerical experience, relatively small grids ($N\approx 5$) are sufficient to reach convergence (i.e., no observable change in the results by further increasing the value of $N.$) The inverses of the overlap matrices can conveniently be calculated by numerical means, and the time derivative of the coefficients $c_j^{\pm}$ are given by Eqs.~(\ref{coeffdyn1})
and (\ref{coeffdyn2}). Using an appropriate integration routine (preferably with adaptive stepsize), the norm
\begin{align}
\langle \Psi|\Psi\rangle(t)=&\sum_{km} (c^+_k(t))^* \mathcal{N}_{km}^{++}(t) c^+_m(t) \nonumber \\ +&\sum_{km} (c^-_k(t))^* \mathcal{N}_{km}^{--}(t) c^-_m(t)
\label{norm}
\end{align}
can be kept close to unity.
This method is particularly efficient when the mean photon number is large. Practically, for $|\alpha|> 10,$ working in photon number eigenstate  basis [see Eqs.~(\ref{ndyn})] becomes increasingly difficult, partly because of the increase of the number of the basis elements that have to be taken into account, but also because of the numerical difficulty of handling the factorials that appear in the expansion coefficients $\langle n|\alpha\rangle.$ Besides the numerical efficiency, using the von Neumann lattice has an additional benefit for the visualization of the results on the phase space: as we shall see in the next subsection, the Wigner function of the mode can be calculated analytically in this basis.

\subsection{Wigner functions}
Since its introduction in 1932 \cite{W32}, the Wigner function became the central tool for describing quantum mechanics on the phase space, allowing for the investigation the connections and most important differences between quantum  mechanics and classical statistical mechanics. Later on, especially in the context of quantum optics, additional quasidistribution functions were introduced \cite{Husimi40,Sudarshan63,Glauber63}, that can also describe the quantum state of light. A detailed summary of these distributions together with the discussion of a wide class of problems on the quantum optical phase space can be found in Ref.~\cite{S01}. In the following we show how the von Neumann lattice allows an elegant calculation of the Wigner function.

Let us note that although a joint Wigner function for the atom-field system has already been constructed \cite{CB96}, in the following we use the traditional approach, in which only the field degrees of freedom are involved. For clarity, let us note that disregarding atomic degrees of freedom is most transparently described by a partial trace over atomic states: For a complete atom-field state $|\Psi\rangle$ we construct the density operator $\rho_{af}=|\Psi\rangle\langle\Psi|$ and use the reduced density operator $\rho_f=\mathrm{Tr}_a \rho_{af}=\langle e|\rho_{af}|e\rangle+\langle g|\rho_{af}|g\rangle$ for the calculation of the Wigner function. Since there is no additional reason for using density operators in the following (there are no mixed states involved), below we use a definition that is based on pure states.

The Wigner function corresponding to a state $|\Psi\rangle$ is essentially the two-dimensional Fourier transform of the characteristic function $\chi$:
\begin{equation}
W_\Psi(\alpha)=\frac{1}{\pi^2}  \int  \!\! \int \chi_\Psi(\alpha) e^{\alpha^*\alpha-\alpha\alpha^*} d^2 \alpha,
\label{W}
\end{equation}
where $\chi_\Psi (\alpha)$ is the expectation value of the displacement operator $D(\alpha)$:
\begin{equation}
\chi_\Psi (\alpha)=\langle \Psi|D(\alpha)|\Psi \rangle.
\label{chi}
\end{equation}
For the case of $t=0,$ we can substitute the form of $|\Psi \rangle$ given by Eq.~(\ref{initial}) into the equation above. By using the orthogonality $\langle-|+\rangle=0,$ we can write:
\begin{align}
&W_\Psi(\alpha)=\\
=&\frac{1}{\pi^2}  \int \!\! \int \sum_{km} (c_k^+)^*c_m^+ \langle\alpha^{(k)}|D(\alpha)|\alpha^{(m)}\rangle e^{\alpha^*\alpha-\alpha\alpha^*} d^2 \alpha \nonumber \\
+&\frac{1}{\pi^2}  \int\!\! \int \sum_{km} (c_k^-)^*c_m^- \langle\alpha^{(k)}|D(\alpha)|\alpha^{(m)}\rangle e^{\alpha^*\alpha-\alpha\alpha^*} d^2 \alpha.
\label{W2}
\end{align}
Note that by using the reduced density operator, one would obtain the same result. (This is due to the fact that the displacement operator $D(\alpha)$ acts as the identity on the atomic subspace.)

It is practical to apply the Baker-Campbell-Haussdorf identity to factorize the exponent in the displacement operator and obtain
\begin{equation}
\langle\alpha^{(k)}|D(\alpha)|\alpha^{(m)}\rangle=e^{-|\alpha|^2/2}e^{\alpha(\alpha^{(k)})^*-\alpha^*\alpha^{(m)}} \langle\alpha^{(k)}|\alpha^{(m)}\rangle.
\end{equation}
By explicitly using the real and imaginary parts of $\alpha ,$ the result above tells us that the Fourier transform of the summands in Eq.~(\ref{W2}) can be calculated analytically (since we are dealing with shifted Gaussians). The generalization of the steps above to the case of time dependent von Neumann lattice coherent states is straightforward, and the final result is given by:
\begin{align}
&W_\Psi(\alpha)=\\
&=\frac{2}{\pi}   \sum_{km}  (c_k^+)^* \mathcal{N}_{km}^{++} c_m^+ e^{-2(\mathrm{Im}\ \alpha-z_{1}^{km})^2} e^{-2(\mathrm{Re}\ \alpha-z_{2}^{km})^2}\nonumber \\
+& \frac{2}{\pi}\sum_{km} (c_k^-)^* \mathcal{N}_{km}^{--} c_m^- e^{-2(\mathrm{Im}\ \alpha-z_{1}^{km})^2} e^{-2(\mathrm{Re}\ \alpha-z_{2}^{km})^2},
\label{W3}
\end{align}
where the explicit notation of the time dependences of the overlap matrices and the coefficients are omitted, and the complex numbers in the exponents are given by
\begin{align}
z_{1}^{km}&=\frac{1}{2}\left(i\mathrm{Re}\ \alpha^{(k)}+\mathrm{Im}\ \alpha^{(k)}-i\mathrm{Re}\ \alpha^{(m)} + \mathrm{Im}\ \alpha^{(m)}\right),  \\
z_{2}^{km}&=\frac{1}{2}\left(-i\mathrm{Im}\ \alpha^{(k)}+\mathrm{Re}\ \alpha^{(k)}+i\mathrm{Im}\ \alpha^{(m)} + \mathrm{Re}\ \alpha^{(m)} \right).
\label{z12}
\end{align}
Note that for the special case when $|\Psi \rangle$ corresponds to a single lattice point, e.g., $|\Psi \rangle=|+\rangle|\alpha^{(k)}\rangle,$ $W_\Psi(\alpha)=2/\pi \exp(-|\alpha-\alpha^{(k)}|^2),$ i.e., the well-known Wigner function of a coherent state $|\alpha^{(k)}\rangle$ is recovered.

As we have seen in this subsection, having determined the coefficients $c_k^{\pm}(t),$ the Wigner function of the state $|\Psi \rangle(t)$ can be calculated analytically, without the need of fast Fourier transform or any other numerical method.

\section{HHG on phase space}
\label{mainsec}

Before focusing on the physical consequences of the model outlined before, it is worth summarizing earlier results that are related. Interaction of a two-level system with a single quantized mode mostly discussed in the framework of rotating wave approximation (RWA) \cite{WM94}, i.e., in the case when the interaction part of the Hamiltonian $H_{af}$ is replaced by $H_{af}^{RWA}=\hbar\Omega(\sigma_+a+\sigma_-a^\dagger),$ where $\sigma_+|g\rangle=|e\rangle, \sigma_+|e\rangle=0$ and $\sigma_-=\sigma_+^\dagger.$ When the initial state of the field is a coherent state, $|\Psi\rangle(0)=|g\rangle|\alpha\rangle$, Rabi oscillations with different frequencies dephase on a time scale that is proportional to $1/\Omega$ (collapse) and -- because of the quantized nature of the radiation field -- they rephase again (revival). The characteristic time of the revival process depends on the mean photon number as well, it is $|\alpha|$ times longer than that of the collapse \cite{ENS80}. These processes has been discussed on phase-space in detail in Ref.~\cite{ER91}. As it was shown, the initially localized (Gaussian) phase-space bump that corresponds to $|\alpha\rangle,$ falls into parts during the time evolution (collapse) and the parts meet again at the revival time. In the strict sense, these results are valid only for the case of a Hamiltonian with RWA, but the time scales for the processes can serve as rough estimations also for the more general case without RWA. For typical experimental situations, e.g., HHG on gas samples when the driving is in the infrared region, the Rabi frequency is by orders of magnitudes below $\omega.$ However, high harmonics are generated during a few, or a few times ten optical cycles only. Therefore collapse and revival are assumed to play minor role on the time scale of HHG. This is why a limited number of basis states centered around the initial coherent state is sufficient to describe the phenomenon.

\bigskip

In the following we focus on the process of HHG as represented on phase space. As one can check easily, RWA does not allow the appearance of high harmonics, that is why we used the Hamiltonian (\ref{HamS}). As a systematic investigation shows, all the terms of this Hamiltonian play significant role in the process. For the sake of completeness, let us recall the case when the field is free, i.e., it does not interact with the atom. As it is known, in this case the time evolution of an initial coherent state $|\alpha\rangle$ is given by $|\alpha e^{-i\omega t}\rangle.$ The corresponding Wigner function will be a Gaussian that circulates clockwise (with a circle time of $T=2\pi/\omega$) along a circle of radius $|\alpha|.$

As it was shown in Sec.~\ref{modelsec}, in the strong field approximation -- when $H_a$ is omitted and the time evolution is governed by $\tilde{H}$) -- the initial state $|\Psi\rangle(0)=|g\rangle|\alpha\rangle=1/\sqrt{2}(|+\rangle+|-\rangle|\alpha\rangle$ splits into two parts, see Eq.~(\ref{coh1t}). Note that this splitting is completely different from the one reported in Ref.~\cite{ER91}: it appears in every optical cycle, much before observable collapse appears. It is important to stress here that $\tilde{H}$ commutes with the (dimensionless) dipole moment operator, $\sigma_x,$ which means that the expectation value of the dipole moment is constant in this case and no HH are generated when the atomic Hamiltonian $H_a$ is omitted.

\begin{figure}[htb]
\includegraphics[width=8cm]{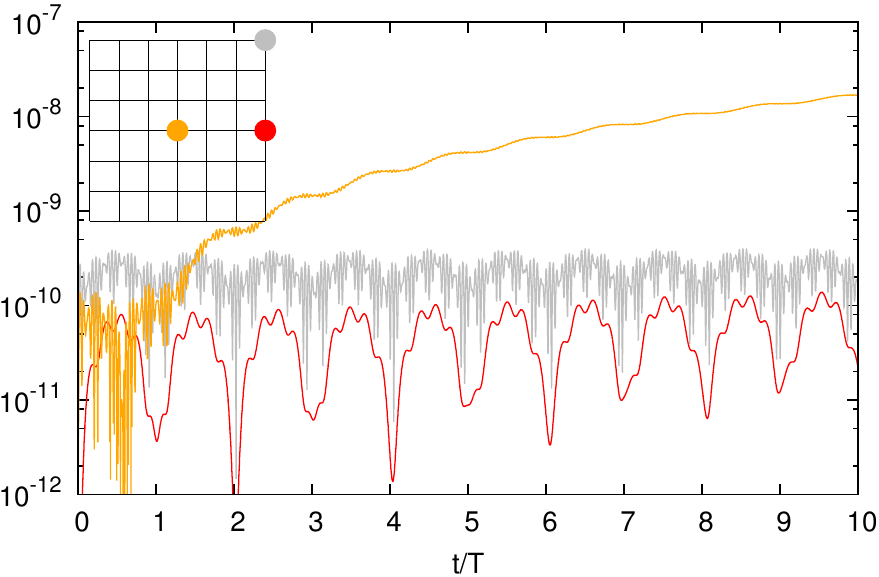}
\caption{The figure corresponds to the complete time evolution with $|\gamma|=0.001.$ Red and grey curves show the weights $|c^+_k|^2+|c^-_k|^2$ of states whose index $k$ correspond to positions on the von Neumann lattice as it is shown by the dots in the inset. The initial state, $|g\rangle |\alpha_0\rangle=100\sqrt{\pi}$ is a coherent state at the center of the lattice (orange dot in the inset), and the orange curve shows $1-|c^+_{k_0}|^2-|c^-_{k_0}|^2$, indicating that Eqs.~(\ref{apmsimple}) mean a good approximation for few optical cycles. }
\label{linefig}%
\end{figure}

Finally let us consider the complete time evolution, which is governed by the full Hamiltonian $H$ given by Eq.~(\ref{HamS}). The question is to what extent the picture we have just discussed changes by the presence of $H_a$. It is clear that $|\Psi\rangle(t)$ is not equal all the time to the superposition of $|+\rangle|\alpha\rangle(t)$ and $|-\rangle|\alpha\rangle(t),$ but -- in the strong field limit -- we expect little deviation from this solution. Numerical calculations with realistic parameters verify this assumption, at least in the sense that state of the mode stays localized on phase space during a few optical cycles. As the red and grey solid lines in Fig.~\ref{linefig}
show, already for a lattice with $N=3,$ the populations of the states at the edges of the grid (furthest away from the localized part) are very low.

\begin{figure}[htb]
\includegraphics[width=8cm]{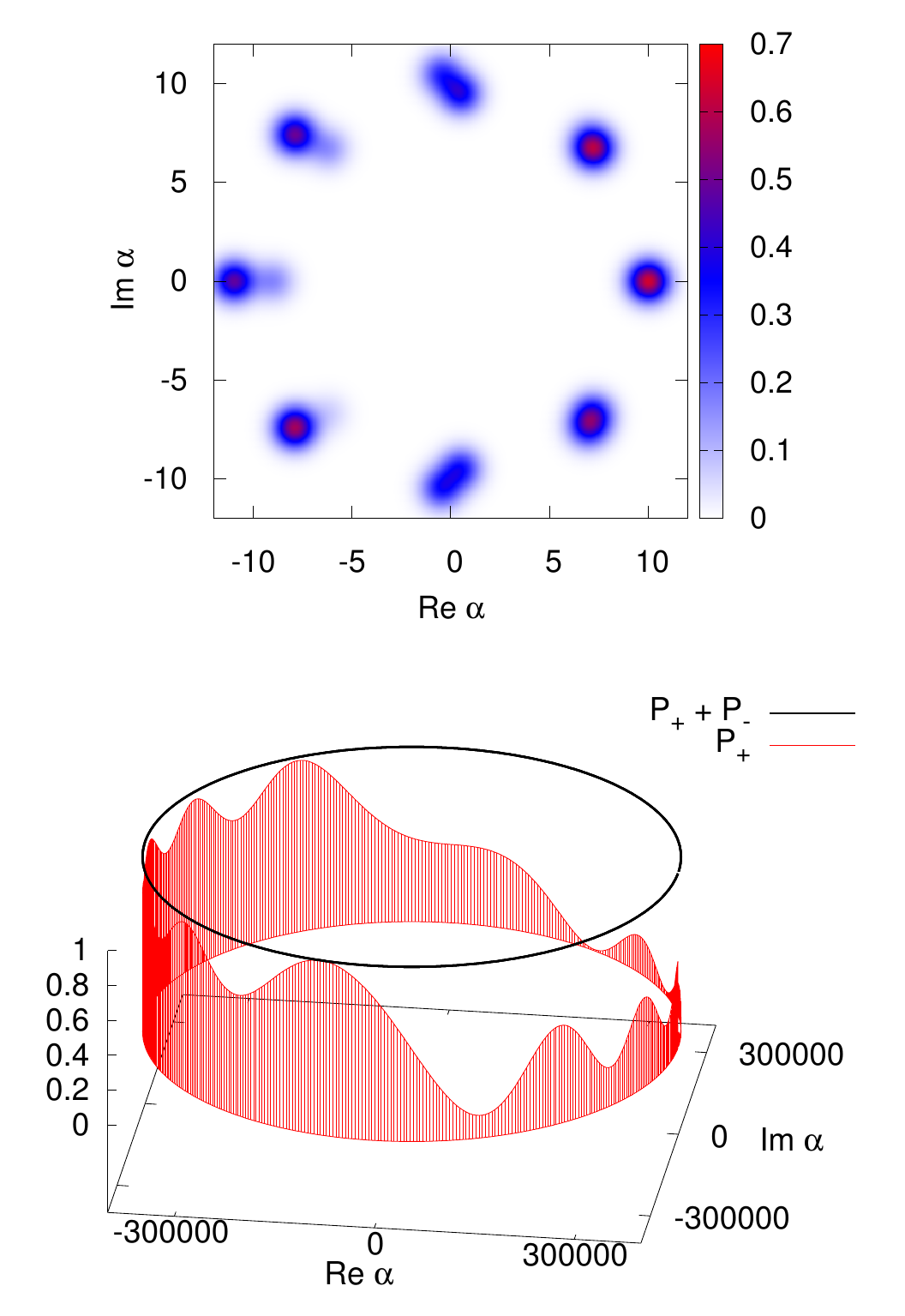}
\caption{Wigner functions visualizing the time evolution induced by the complete Hamiltonian. Top panel: the eight different localized bumps correspond to $W(\alpha,t)$ at $t=0, 1/8T, 2/8T, \ldots, 7/8T$ (first optical cycle). The parameters for this panel are $|\alpha_0\rangle=10,$ $\gamma=0.5.$ The bottom panel corresponds to the more realistic values of $|\alpha_0\rangle=4\times10^5,$ $\gamma=10^{-5}$ and shows the probability $P_+$ that system is in the state $|+\rangle|\alpha\rangle(t)$ along the corresponding phase space trajectory for the first optical cycle. The black curve on the top of this panel shows $P_+ + P_-,$ which, as we can see, is very close to unity.}
\label{Wdynfig}%
\end{figure}

It is instructive to investigate the minimal case when we estimate the state of the system by
\begin{equation}
|\Psi\rangle(t)=\frac{1}{\sqrt{2}}\left[c^-(t)|-\rangle|\alpha\rangle(t) + c^+(t)|+\rangle|\alpha\rangle(t) \right],
\end{equation}
where the time evolution of the states appearing in the right hand side is given by Eq.~(\ref{coh1t}). Using this and the action of $H_a$ on the states $|\pm\rangle$, we obtain:
\begin{align}
i\hbar \dot{c}^{+}&= c^{-} \langle \alpha_+(t)|\alpha_-(t)\rangle e^{i[\delta_-(t)-\delta_+(t)]},  \nonumber \\
i\hbar \dot{c}^{-}&= c^{+} \langle \alpha_-(t)|\alpha_+(t)\rangle e^{-i[\delta_-(t)-\delta_+(t)]}.
\label{apmsimple}
\end{align}
Since $|\gamma|\ll 1,$ the inner product $\langle \alpha_+(t)|\alpha_-(t)\rangle$ is close to unity. Additionally, the exponential terms on the right hand side turns out to have the time dependence of $\exp(\pm \alpha \gamma \sin \omega t),$ which is typical for high harmonic generation. It is important to note that the crucial parameter here is $|\alpha\gamma|=\alpha\Omega/\omega.$ That is, the larger $\alpha$ is (the more photons the exciting mode has) and the larger the ratio $\Omega/\omega$ is, the more harmonics will be generated. Although this result is based on a strong approximation (the validity of which is shown by the orange line in Fig.~\ref{linefig}), it qualitatively holds also in the exact case.

Fig.~\ref{Wdynfig} summarizes the results discussed so far. In the top panel the Wigner function is shown at different time instants for parameters that are ideal for the visualization of the dynamics. The mean photon number is considerably larger for the bottom panel, where the probability $P_+$ of finding the system in the state $|+\rangle|\alpha\rangle(t)$ is shown. As we can see, although the atomic Hamiltonian $H_a$ induces fast transitions between the states $|+\rangle|\alpha\rangle(t)$ and $|-\rangle|\alpha\rangle(t),$ at the beginning of the time evolution the system remains in a superposition of these two states to a very good approximation.

\section{HHG spectra}
\label{spectrumsec}
Besides the phase-space picture presented in the previous section, it is also important to see HHG spectra, which are probably the most distinctive signatures of the process. To this end, we calculate the expectation value of the operator $\sigma_x$ (that is proportional to the dipole moment) in the time dependent solution of the problem as given by Eq.~(\ref{timedep}). Since $\sigma_x|\pm\rangle=\pm|\pm\rangle,$ we have:
\begin{align}
\langle\sigma_x\rangle(t)=&\sum_{j,k} \left[c^+_k(t)\right]^* c^+_j(t) \mathcal{N}_{jk}^{++}(t) \\
-&\sum_{j,k} \left[c^-_k(t)\right]^* c^-_j(t) \mathcal{N}_{jk}^{--}(t).
\end{align}
Once we obtained the coefficients $c^\pm_j$ and the overlap matrices $\mathcal{N}_{jk}^{\pm\pm}$ as functions of time, the expectation value above can readily be calculated, and its power spectrum provides the HHG spectrum that corresponds to the same initial conditions as the solution $|\Psi\rangle(t)$.

\begin{figure}[htb]
\includegraphics[width=8cm]{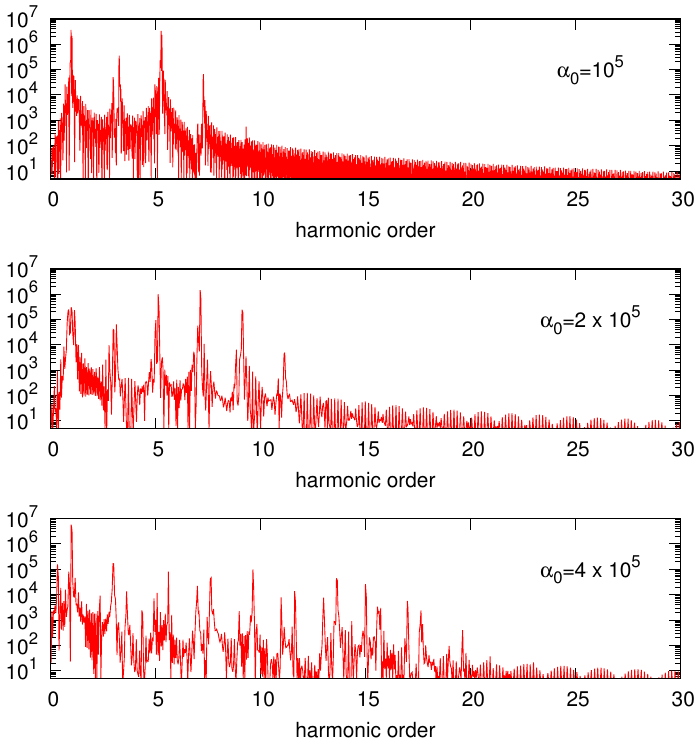}
\caption{HHG spectra for different initial states $|\Psi\rangle(0)=|g\rangle|\alpha_0\rangle,$ where $\alpha_0$ is indicated in the subfigures, and $\gamma=10^{-5}.$ The Fourier transform of $\langle\sigma_x\rangle(t)$ were calculated using an interval of 20 optical cycles.}
\label{spectrfig}%
\end{figure}

Fig.~\ref{spectrfig} shows HHG spectra for different exciting intensities as determined by the index of the initial coherent state in $|\Psi\rangle(0)=|g\rangle|\alpha_0\rangle.$ The usual qualitative features of high harmonic spectra are visible: i) we have pronounced peaks, ii) a plateau region and also iii) a cutoff frequency, which increases for stronger excitations. Unlike these general properties of the spectra, the internal structure of the peaks is specific to the case of the two-level system. In fact, the description of the exciting field also affects these details: comparing the model presented here and the case of classical, time-dependent driving, we have seen that the general features i)-iii) are qualitatively the same (including the dependence of the cutoff on the intensity). On the other hand, the detailed structure of the HH peaks are different, in the classical case we usually see twin peaks which can be understood using  Floquet's theorem \cite{GC16}, while these maxima in the spectra are more structured in our case. Since our aim is to provide a clear, instructive phase space picture, the detailed analysis of this difference is beyond the scope of the current paper.

\section{Generalization}
\label{generalizationsec}
So far we considered  the case of a two-level system that interacts with a quantized mode of electromagnetic radiation that is initially in a coherent state with a high mean photon number. In this section we discuss how more general material systems and more complex fields can be described. This helps distinguishing between general and model-specific properties of the physical picture that was delineated so far. At some points, the treatment of this section will be qualitative only, since, instead of describing specific detailed properties of various physical systems, our intention is to focus on general questions.

Let us first replace the two-level atom by an arbitrary material system. This means considering
\begin{equation}
H'=H'_a+H_f+H'_{af}=H'_a +\hbar\omega a^\dagger a + \hbar\Omega \mathcal{D}(a+a^\dagger),
\label{Hamp}
\end{equation}
where $H'_a$ and the coupling operator $\mathcal{D}$ (which, for the sake of definiteness, will be called dipole moment operator) need not be specified in more details. However, in order to simplify the technical details, it is worth assuming that they act on a finite dimensional Hilbert-space, and, consequently, their spectra are discrete. We will explicitly use only the eigenstates and eigenvalues of the dipole moment operator: $\mathcal{D}|k\rangle=d_k |k\rangle,$ where $\{d_k\}_{k=1}^M$ are real numbers. (The states $|k\rangle$ will be playing the role of $|+\rangle$ and $|-\rangle$.) As we mentioned after Eq.~(\ref{transformedH}), all the discussion that led to Eq.~(\ref{transformedH}) are valid also with the replacement of $\sigma_x$ by the more general operator $\mathcal{D}.$ That is, once the eigenvalue problem of $\mathcal{D}$ is solved, $\tilde{H}'=H_f+H'_{af}$ can be diagonalized, similarly to the case of Sec.~\ref{modelsec}. However, in this case the atomic part of the diagonalized state cannot be arbitrary, since generally $\mathcal{D}^2$ is not proportional to unity. That it, the eigenstates $\tilde{H}'$ should also be one of the eigenstates of $\mathcal{D},$ e.g. $|k\rangle$, when the corresponding eigenenergy will contain a term proportional to $d_k^2,$ see the next paragraph.

Next, let us consider the interaction with a multimode field, i.e.,
\begin{equation}
H''=H''_a+H''_f+H''_{af}=H''_a +\hbar\sum_{i=1}^{M}\omega_i a_i^\dagger a_i + \hbar\Omega_i \mathcal{D} (a_i+a_i^\dagger),
\label{Hampp}
\end{equation}
where the coupling strength of the interaction can depend on $\omega_i,$ that is being taken into account by the index $i$ of the Rabi frequencies $\Omega_i$. As we can see, in this case the Hamiltonian $\tilde{H}''=H''-H_a''$ factorizes:
\begin{equation}
\tilde{H}''=\sum_i \tilde{H}''_i=\hbar\sum_i\omega_i a_i^\dagger a_i + \hbar\Omega_i \mathcal{D} (a_i+a_i^\dagger),
\label{Htildepp}
\end{equation}
and $\left[\tilde{H}''_i, \tilde{H}''_j\right]=0.$
This means that in the "strong-field approximation," i.e., when the atomic Hamiltonian can be be neglected when compared to the sum of the free-field Hamiltonian and the interaction term, different modes of the quantized fields are independent. Note that each operator $\tilde{H}''_i$ in the sum above -- apart from the index $i$ -- is the same as $\tilde{H}'$ that was introduced earlier in this section. By using the multimode displacement operator $\tilde{D}(\{-\Omega_i/\omega_i\})=\tilde{D}(-\Omega_1/\omega_1)\otimes\ldots \otimes\tilde{D}(-\Omega_M/\omega_M)$ (where we retained explicit tensorial notation for clarity), we can see that tensor products of displaced number states
\begin{equation}
|k\rangle\otimes\tilde{D}(-\Omega_1/\omega_1) \left|n_1\right\rangle\otimes\cdots\otimes \tilde{D}(-\Omega_M/\omega_M) \left|n_M\right\rangle
\end{equation}
are eigenstates of $\tilde{H}''$ with the eigenvalues of $\sum_i \hbar \omega_i(n_i-d_k^2 \Omega_i^2/\omega_i^2).$ Additionally, the time evolution of a multimode coherent state $|k\rangle\otimes|\{\alpha_i\}\rangle=|k\rangle\otimes|\alpha_1\rangle\otimes \cdots\otimes|\alpha_M\rangle$ as generated by $\tilde{H}''$ can also be calculated as a straightforward generalization of Eq.~(\ref{coh1t}).

\bigskip

That is, the general qualitative picture of the process of high-order harmonic generation on phase space can be describes as follows: The dominant part of the complete Hamiltonian ($H,$ for the sake of simplicity without any prime) is the one describing the free field and its interaction with matter. The time evolution generated by these terms (strong-field approximation) can be solved analytically. The Gaussian Wigner functions that correspond to the initial coherent states of the quantized modes fall apart and form as many Gaussian peaks as the number of the eigenstates of the dipole moment operator is needed to expand the initial state. At the end of the optical cycle these separate Gaussians merge again, and the process gets repeated periodically. The separation of these different Gaussian peaks is determined by the strength of the light-matter interaction.

The atomic Hamiltonian, $H_a,$  is the weakest part of the complete Hamiltonian $H$ (e.g., its expectation value is much less than that of $\tilde{H}=H-H_a$). However, its presence is necessary for the generation of the harmonics (in fact, any radiation), since the dipole moment (expectation value) is constant otherwise. Besides generating secondary radiation, the fact that $H_a$ does not commute with $\tilde{H}$ leads to transitions between the states that correspond to different Gaussian parts of the  Wigner function that evolve independently for $H_a=0.$ Moreover, the presence of $H_a$ increases the size of the the phase-space regions on which the corresponding Wigner functions are considerably different from zero for the various modes. However, this broadening of the Wigner functions  means a weak effect on the timescale of HHG, thus using finite parts of the von Neumann lattices for each mode means an efficient numerical approach also in the general cases.

\section{Summary}
\label{finalsec}
The description of high-order harmonic generation on quantum optical phase space was considered. For the sake of simplicity, we discussed finite dimensional systems and have shown that in the strong-field limit (i.e., when the Hamiltonian describing the material system can be omitted) the analytic solution of the dynamics allows for clear phase space interpretation. The corresponding Wigner functions were determined using the von Neumann lattice. The properties of the HHG process allowed us to develop an efficient numerical method that can solve the complete dynamics for arbitrarily large photon numbers. The role of different terms in the Hamiltonian were investigated systematically, and we saw that all of them is needed for the appearance of high harmonics.

\section*{Acknowledgments}
We thank P. Tzallas and M. G. Benedict for useful discussions.
This research was performed in the framework of the project Nr. GINOP-2.3.2-15-2016-00036 titled Development and application of multimodal optical nanoscopy methods in life and material sciences.
The project has been also supported by the European Union, co-financed by the European Social Fund, Grant No. EFOP-3.6.2-16-2017-00005.
Partial support by the ELI-ALPS project is also acknowledged.
The ELI-ALPS project (GINOP-2.3.6-15-2015-00001) is supported
by the European Union and co-financed by the European Regional Development Fund.
Our work was also supported by the grant TUDFO/47138-1/2019-ITM FIKP of the Ministry of Innovation and Technology, Hungary.


\end{document}